\documentclass[%
 reprint,
 amsmath,amssymb,
 aps,
]{revtex4-2}

\usepackage{graphicx}
\usepackage{dcolumn}
\usepackage{bm}
\usepackage{textgreek}
\usepackage{amsmath}
\usepackage{color,soul}
\usepackage{geometry}
\usepackage[utf8]{inputenc}
\usepackage{graphicx}
\usepackage{dcolumn}
\usepackage{amsmath}
\usepackage{graphicx}
\usepackage[dvipsnames]{xcolor}
\usepackage{hyperref}
\usepackage[sort&compress]{natbib}

\begin{document}

\preprint{APS/123-QED}

\title{Ultrafast laser welding of silicon}

\author{Maxime Chambonneau,$^{1}$ Qingfeng Li,$^{1}$ Markus Blothe,$^{1}$ Stefan Nolte$^{1,2}$}
\affiliation{$^{1}$Institute of Applied Physics, Abbe Center of Photonics, Friedrich Schiller University Jena, Albert-Einstein-Straße 15, 07745 Jena, Germany.}
\affiliation{$^{2}$Fraunhofer Institute for Applied Optics and Precision Engineering IOF, Center of Excellence in Photonics, Albert-Einstein-Straße 7, 07745 Jena, Germany}

\date{\today}

\begin{abstract}

While ultrafast laser welding is an appealing technique for bonding transparent workpieces, it is not applicable for joining silicon samples due to nonlinear propagation effects which dramatically diminishes the possible energy deposition at the interface. We demonstrate that these limitations can be circumvented by local absorption enhancement at the interface thanks to metallic nanolayer deposition. By combining the resulting exalted absorption with filament relocation during ultrafast laser irradiation, silicon samples can be efficiently joined. Shear joining strengths $>4$~MPa are obtained for 21-nm gold nanolayers without laser-induced alteration of the transmittance. Such remarkable strength values hold promises for applications in microelectronics, optics, and astronomy.

\end{abstract}

\maketitle

\section{Introduction}

Ultrafast laser welding increasingly attracts attention as it offers the possibility to join a wide variety of materials in a fast, clean, and adhesive-free way \cite{Richter2016}. It relies on the propagation of ultrashort laser pulses through a first workpiece, energy deposition at the interface with a second workpiece, which locally leads to melting, resolidification, and the creation of strong bonds between the two workpieces. This technique can be applied to similar and dissimilar materials \cite{Carter2014}. However, until recently, ultrafast laser welding was limited to configurations where the first workpiece is transparent and exhibits a large band gap (e.g., glass, polymer) \cite{Mingareev2012,Carter2014,Zhang2015,Richter2016,Zhang2018,Penilla2019,Cvecek2019,Sahoo2020}. In contrast, silicon as a narrow band-gap material exhibits a high nonlinear refractive index causing the formation of filaments \cite{Zavedeev2016,Kononenko2016a,Fedorov2016,Chambonneau2021a}. Consequently, in such a nonlinear propagation regime, the energy deposition is strongly limited and delocalized, which is a major hurdle when one aims at joining silicon to another material. Recently, we have demonstrated that silicon--metal ultrafast laser welding is achievable when the nonlinear focal shift is precompensated, i.e., the maximum fluence is relocated at the interface between both materials \cite{Chambonneau2021}. In this configuration, the free electrons of the metal linearly absorb the laser flux by inverse Bremsstrahlung. However, silicon--silicon ultrafast laser welding is still an open issue, as the absorption is necessarily nonlinear (multi-photon ionization)---and thus, the energy deposition at the interface between two silicon samples is less efficient than in the silicon--metal configuration. While it was shown that nanosecond pulses \cite{Sopena2022a} or continuous light \cite{Sari2008} can be employed for joining silicon samples, thermal effects in such regimes are expected to decrease the degree of control on the energy deposition. The demonstration of an ultrafast laser-based technique for joining stacked semiconductors such as silicon would have direct applications in microelectronics, and even beyond. For instance, this technique would have strong potential for fabricating optical components based on stacked semiconductors with applications in laser manufacturing \cite{Keller1996}, and astronomy \cite{Zhang2019}.\\

In this paper, we demonstrate ultrafast laser welding of silicon. To optimize the energy deposition at the interface between the silicon samples, our recently developed filament relocation technique is applied \cite{Chambonneau2021}. Prior to the welding, the absorption at the interface is increased by the deposition of a gold nanolayer. Welding attempts without nanolayer failed, as the interface is not modified with the laser radiation. The influence of the gold nanolayer thickness on the transmittance of the joined samples and the bonding strength is evaluated. The best performance is obtained for the thinnest gold nanolayers (21~nm), which yield shear joining strengths $>4$~MPa between the samples. This result originates from the high temperatures which can be reached for the thinnest nanolayers, as confirmed by the increase in the laser-induced damage threshold (LIDT) with the nanolayer thickness. Ultimately, observations of the welds after separation show that the thickest nanolayers lead to inefficient joining between the silicon samples.

\section{Rationale}

In order to devise a procedure for performing ultrafast laser welding of silicon, let us develop in this Section our reasoning based on the existing state of the art. First, let us recall that the only demonstrations of laser-joined silicon samples have been realized with nanosecond pulses \cite{Sopena2022a} and continuous light \cite{Sari2008}. This is consistent with the fact that this long-interaction regime is adapted to the production of repeatable modifications inside silicon \cite{Verburg2014,Chambonneau2016,Tokel2017,Chambonneau2018b,Wang2019a,Turnali2019,Wang2021}. However, the modifications produced in this regime exhibit large heat-affected zones \cite{Chichkov1996}, as a consequence of thermal runaway during the interaction. Such large-sized modifications are likely to severely alter the material properties, causing for instance degradation of the thermal and electrical properties, wavefront distortions, light scattering, and tilt between the samples.\\

In contrast, ultrashort pulses ($<10$~ps) lead to limited heat-affected zone. However, permanent modifications are extremely challenging to produce in the bulk and the exit surface of silicon with ultrashort laser pulses \cite{Chambonneau2021a}, unless complex approaches are implemented for optimizing the spatial \cite{Sreenivas2012,Chanal2017,Blothe2022}, spectral \cite{Richter2020,Mareev2022}, or temporal \cite{Mori2015a,Wang2020a,Chambonneau2021b} properties of the irradiation. These limitations originate from the high nonlinear refractive index of silicon (on the order of $10^{-14}$~cm$^{2}$~W$^{-1}$ \cite{Bristow2007,Lin2007}), which causes micro-filamentation at modest peak powers---and thus, an intensity clamping similar to that described for transparent media \cite{Couairon2007,Berge2007}. The main difference between silicon and transparent media is that the intensity saturation level in silicon is below the damage threshold.\\

In the context of ultrafast laser welding, the challenge to produce permanent modifications inside silicon leads to a paradoxical situation. Traditionally, for glasses, higher bonding strengths are observed for samples kept in optical contact during the laser welding \cite{Richter2015,Tan2017}. However, by definition, two samples in optical contact behave optically as a single sample. For silicon, this implies that attempting to join samples in optical contact practically means to modify the bulk of a single silicon sample, which is extremely hard to achieve. On the other hand, when the silicon samples are not in optical contact, the air gap between these would result in high losses ($\approx 30\%$ at each air--silicon interface) due to the large refractive index mismatch between silicon and air. Furthermore, as an excellent thermal insulator, air would hinder efficient transfer of the laser-produced heat, as well as material interpenetration from one sample to the other---and thus, the creation of strong bonds between the samples.\\

One way to sidestep this paradox is to engineer the exit surface so that the energy deposition is locally increased. For instance, surface grinding causes increased roughness, in turn acting as damage precursor through electric field enhancement. While this approach is suitable for surface-seeded inscription of waveguides in the bulk of silicon \cite{Alberucci2020}, optical contact between rough samples cannot be achieved. In order to reach the damage threshold on the exit surface of silicon, and simultaneously preserve low surface roughness, the deposition of a thin metallic film is an excellent option. Indeed, metals exhibit free electrons that are able to linearly absorb the laser flux by inverse Bremsstrahlung, which is particularly efficient for infrared wavelengths such as inevitably employed for reaching the transparency domain of silicon. This exalted absorption on the exit surface is supported by different studies. First, selective laser ablation of gold films deposited on the exit surface of silicon was shown in Refs.~\cite{Lei2016,Astrauskas2021}. Moreover, this approach is consistent with our recent demonstration of silicon--metal ultrafast laser welding \cite{Chambonneau2021}.\\

Various metals can be considered as the absorbing layer between two silicon samples, as most of them exhibit analogous optical properties in the infrared. In our study, we selected gold as it does not oxidize, making the joined samples less prone to aging problems. Finally, one must mention the existing compromise concerning the thickness of the absorbing layer. On the one hand, an extremely thin film would result in insufficient absorption enhancement, and the silicon samples are unlikely to be joined. On the other hand, excessive film thickness would hinder the interpenetration of the silicon samples. A reasonable compromise consists of using film thicknesses on the same order of magnitude as the optical penetration depth of the metal ($\approx 10$~nm for gold). The influence of the gold nanolayer thickness on the bonding quality is examined below.

\section{Experimental arrangement}

The experimental arrangement for performing silicon--silicon ultrafast laser welding is identical to that employed in the silicon--metal configuration as described in Ref.~\cite{Chambonneau2021}. It relies on pulses at a wavelength of 1555~nm delivered at $\Omega_{0}=100$~kHz repetition rate by an Er-doped fiber laser (Raydiance Inc., Smart Light 50). The pulse duration is adjusted to 9.8~ps (full width at half-maximum) to avoid excessive nonlinearities and achieve better welding than with sub-picosecond pulses \cite{Penilla2019}. For the same reasons, the pulse energy is set to 900~nJ, which corresponds to the nonlinear propagation regime where the peak fluence in the bulk of silicon saturates to 125~mJ/cm$^{2}$ without catastrophic focusing and defocusing dynamics (see Section~S1, Supporting Information). The beam is focused with an objective lens of numerical aperture $\text{NA}=0.26$ (Mitutoyo, M Plan Apo NIR $10\times$). In the linear propagation case, the Gaussian beam diameter at the focus is $2w_{0}=5.4$~$\mu$m, as measured with an InGaAs camera and confirmed by simulations with the \textit{InFocus} vectorial model \cite{InFocus2021,Li2021}.\\

The first and second silicon samples (hereafter referred to as the ``top'' and ``bottom'' samples, respectively) are undoped, $<$100$>$-oriented, double-side polished, with resistivity $>200$~$\Omega$~cm, total thickness variation $<10$~$\mu$m, bow $<30$~$\mu$m, and wrap $<30$~$\mu$m. The dimensions of the top and bottom sample are $5\times5\times1$~mm$^{3}$ and $20\times5\times0.5$~mm$^{3}$, respectively. A gold nanolayer is deposited on the entrance surface of the bottom sample by sputtering. The nanolayer thickness is adjusted by changing the sputtering time, and measured with a stylus profiler. Gold nanolayer thicknesses of 0 (i.e., no nanolayer), 21, 28, 43, 70, and 88~nm have been tested. For each configuration, the experiments have been performed on five sample pairs for repeatability. The laser-induced damage threshold (LIDT, in J/cm$^{2}$) is evaluated for each nanolayer by applying 100,000 pulses focused on different sites with the same focusing objective lens as in the welding experiments.\\

\begin{figure*}
\centering
\includegraphics[width=0.75\linewidth]{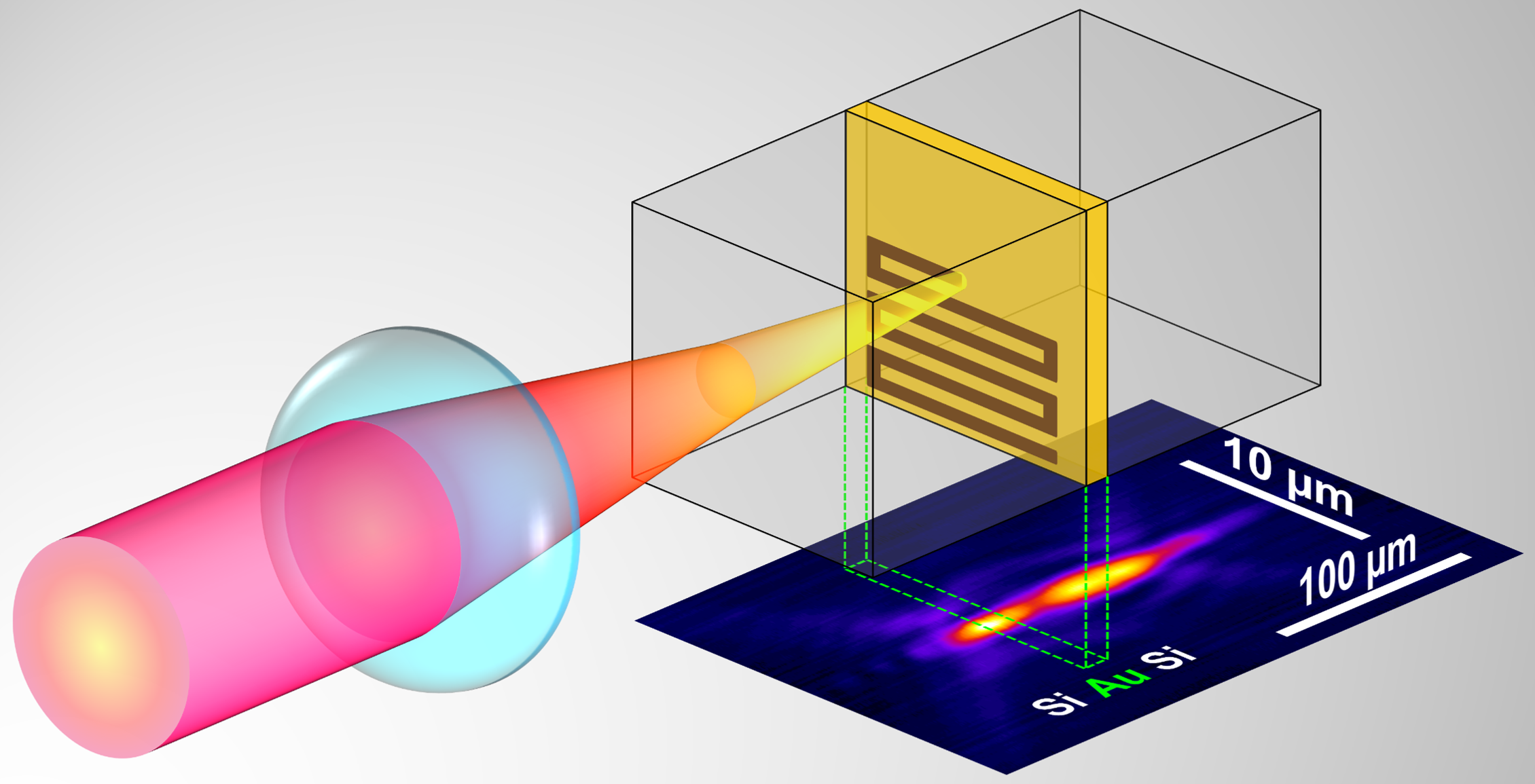}
\caption{\label{fig:FigSetup} Principle of filament relocation for ultrafast laser welding of silicon. A nonlinear propagation image (color map) is recorded. The focusing lens is then moved so that the nonlinear focal shift $\Delta z$ is precompensated prior to the welding, which optimizes the energy deposition at the gold nanolayer (i.e., the interface).}
\end{figure*}

During welding, the samples are maintained in optical contact by means of a clamping system described in Ref.~\cite{Li2022}, so that there is no air gap between them. The welding pattern is a raster scan at a scanning speed $v=1$~mm~s$^{-1}$, with 40-$\mu$m distance between the lines, and a total scanned area of $4.5\times4.5$~mm$^{2}$. The number of pulses per point is $N=2w_{\text{nl}}\Omega_{0}/v=2170$, where $2w_{\text{nl}}=21.7$~$\mu$m is the beam diameter at 1/e$^{2}$ at the nonlinear focus (see Section~S1, Supporting Information), and $\Omega_{0}=100$~kHz is the laser repetition rate. A customized dark-field microscope enables us to accurately position the entrance surface of the top samples with respect to the geometrical focus. Two steps are then applied to optimize the energy deposition at the interface. First, the geometrical focus is positioned at the interface by moving it from the entrance surface of the top sample by $n_{0} \times t_{\text{top}}$ downstream the laser, where $n_{0}=3.48$ is the refractive index of silicon at a wavelength of 1555~nm \cite{Li1980}, and $t_{\text{top}}$ is the thickness of the top sample. In addition, we apply a second step described in Ref.~\cite{Chambonneau2021} and illustrated in Fig.~\ref{fig:FigSetup}, which consists of filament relocation. To do so, the nonlinear focal shift $\Delta z$ is determined with nonlinear propagation imaging. For the input pulse energy used in the welding experiments (900~nJ, corresponding to an input peak power $P_{\text{in}} \approx 81$~kW), $\Delta z=114$~$\mu$m, which is precompensated by moving the objective lens downstream the laser by $\Delta z/n_{0}$. These two steps ensure that the fluence at the interface is optimized.\\

\textit{Ex}-\textit{situ} characterization of the welds is performed by various means listed below.
\begin{itemize}
\item A customized infrared transmission microscope is used for inspecting the interface. It relies on broadband light emission (Thorlabs, QTH10/M), an objective lens (Mitutoyo, M Plan Apo NIR $20\times$), a tube lens (Thorlabs, TTL200-S8), and an InGaAs camera (Xenics, Bobcat 320).
\item Measurements of the total transmittance are carried out with continuous light emitted at a wavelength of 1300~nm by a superluminescent diode light source (Thorlabs, S5FC1018P). The power is monitored before and after light passes through the welds by means of an integrating sphere (Thorlabs, S148C).
\item Breaking tests are performed by applying a force on the edge of the top sample with an indenter (cylindrical shape, 7.4-mm diameter), until it separates from the bottom sample. The corresponding breaking force is recorded by a force gauge (RS PRO, 111-3690) connected to the indenter. The shear joining strength is obtained by calculating the ratio between the measured force and the apparent welded area observed under transmission microscopy. When no laser-modified zones are observed under transmission microscopy, this area is taken as the total scanned area ($4.5 \times 4.5$~mm$^{2}$).
\item Observations of the top sample after breakage are carried out by means of standard bright-field optical microscopy in reflection, as well as Raman spectroscopy.
\end{itemize}

\section{Results and discussion}

\begin{figure}
\centering
\includegraphics[width=0.9\linewidth]{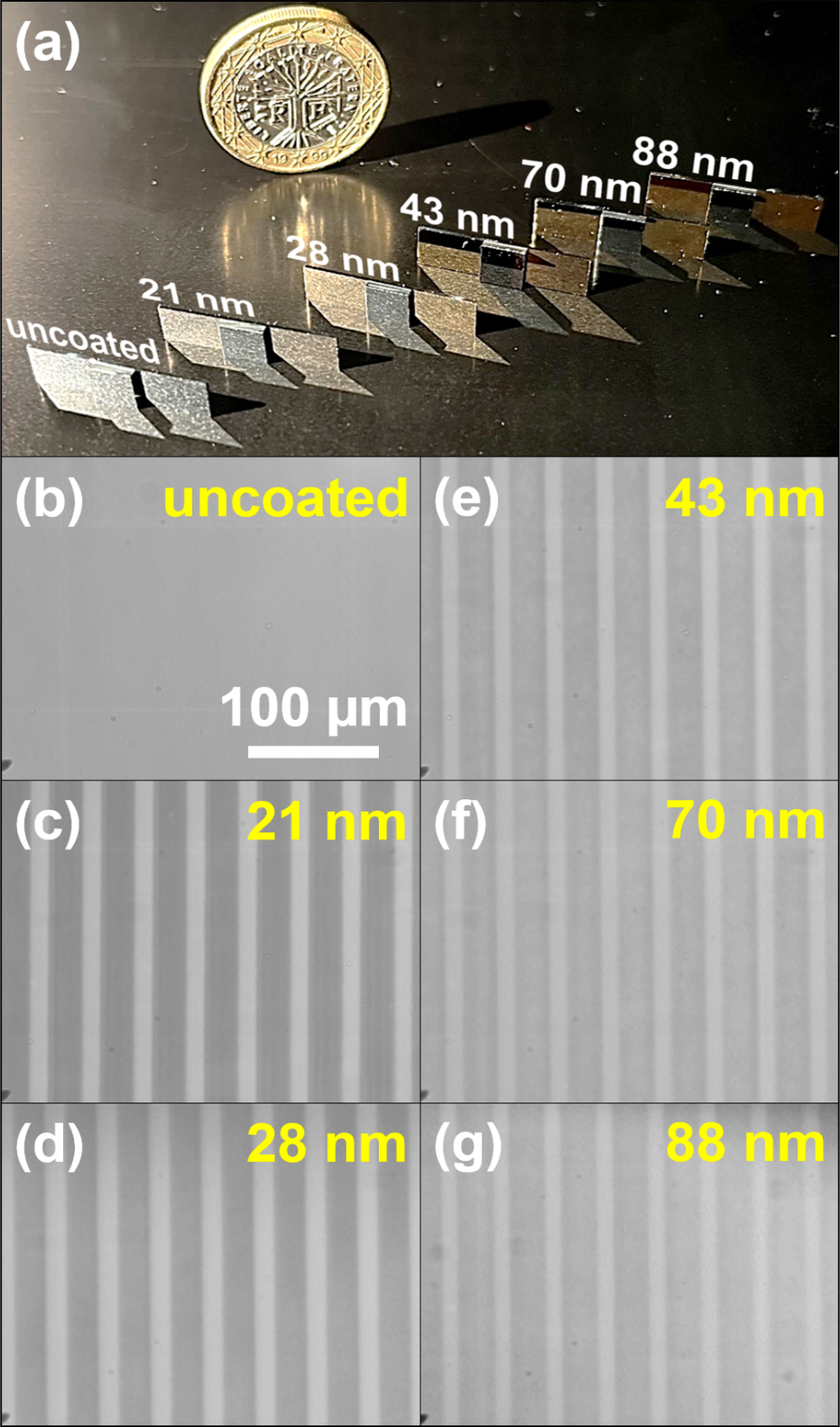}
\caption{\label{fig:FigIRmicroscopy} (a) Photograph of ultrafast laser-welded silicon samples with various gold nanolayer thicknesses. (b)--(g) Corresponding infrared transmission microscopy images of the joined samples at the interface. The dark and bright features correspond to the laser-written tracks and non-irradiated areas, respectively. The spatial scale applies to all images.}
\end{figure}

As a first result illustrated in Fig.~\ref{fig:FigIRmicroscopy}(a), in all configurations, the samples hold together. This may originate from the optical contact between the samples which is preserved after irradiation, or from the laser-produced bonding. Conversely, when the samples are not in optical contact, the sample pair is easily separable by hand, as shown in Section~S2, Supporting Information. Typical infrared transmission microscopy observations of the interface are shown in Figs.~\ref{fig:FigIRmicroscopy}(b)--(g) for different gold nanolayer thicknesses. In Fig.~\ref{fig:FigIRmicroscopy}(b), no noticeable features attributable to laser irradiation are observed when there is no gold nanolayer, showing that no in-volume modifications are produced (this will be further examined below). In contrast, laser-written tracks [dark features in Figs.~\ref{fig:FigIRmicroscopy}(c)--(g)] are observed when a gold nanolayer is deposited on the bottom silicon sample prior to the welding. This result validates our approach as the deposited gold nanolayer exalts the energy deposition at levels above the modification threshold. The average width of the laser-written tracks is $\approx 25.9$~$\mu$m, with a standard deviation of $\approx 1.2$~$\mu$m. This track width value is much larger than the beam diameter ($2w_{0}=5.4$~$\mu$m) because of nonlinear propagation effects, as shown in Section~S1, Supporting Information.\\

\begin{figure}
\centering
\includegraphics[width=\linewidth]{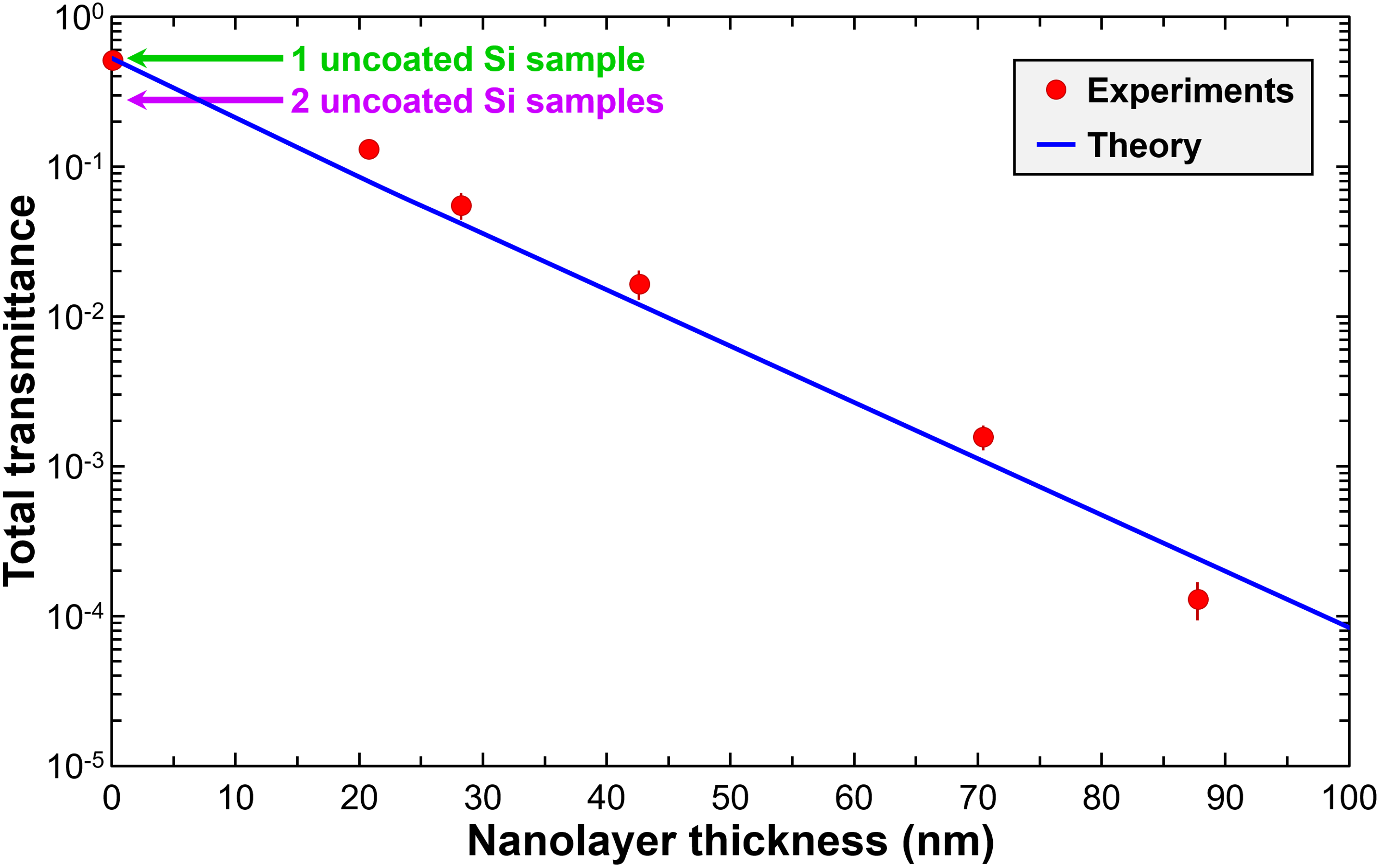}
\caption{\label{fig:FigTransmission} Dependence of the total transmittance of the welds as a function of the gold nanolayer thickness. The red points correspond to experimental measurements with continuous light at a wavelength of 1300~nm. These have been averaged over five welds, and the error bars stand for the standard deviation. The blue curve corresponds to calculations according to Eq.~(\ref{eq:TotTransmJoint}). The green and purple arrows indicate the theoretical transmittance for one, and for two silicon samples which are not in optical contact, respectively.}
\end{figure}

In order to evaluate if the ultrafast laser processing severely alters the optical properties of the samples, the total transmittance has been calculated assuming no decrease caused by the laser-modified zones. To do so, multiple reflections at both air--silicon interfaces, and linear absorption in the gold nanolayer are taken into account with Fresnel equations and Beer--Lambert law, respectively. Under these conditions, as detailed in Section~S3.2, Supporting Information, the total transmittance $T_{\text{tot}}$ of the welds reads
\begin{equation}
\label{eq:TotTransmJoint}
T_{\text{tot}} = \frac{T^{2} e^{\alpha d}}{e^{2\alpha d}-R^{2}},
\end{equation}
where $R=\left( n_{\text{Si}}-1 \right)^{2}/\left( n_{\text{Si}}+1 \right)^{2}$ and $T=1-R$ are the Fresnel coefficients for reflection and transmission, respectively, $\alpha \approx 8.65 \times 10^{7}$~m$^{-1}$ \cite{Yakubovsky2017,Rosenblatt2020} and $d$ are the attenuation coefficient and the thickness of the gold nanolayer, respectively. The corresponding calculations are displayed in Fig.~\ref{fig:FigTransmission} (blue curve). As expected, $T_{\text{tot}}$ decreases exponentially with the nanolayer thickness.\\

These calculations are compared to experimental measurements taken in the irradiated zones (red points). Similar values are obtained when measuring the total transmittance in the non-irradiated zones. Without gold nanolayer, the transmittance of the stack is close to the theoretical value for a single sample (green arrow), which reads $2n_{\text{Si}}/(1+n_{\text{Si}}^{2}) \approx 0.53$, where $n_{\text{Si}}$ is the refractive index of silicon at the considered wavelength [see Section~S3.1, Supporting Information Eq.~(S5)]. In comparison, the theoretical value for the transmittance of two silicon samples with an air gap ($\approx 0.28$, purple arrow) is much lower than the measurements in Fig.~\ref{fig:FigTransmission}, which indicates that the samples are in optical contact. For stacks including gold nanolayers, the experimental measurements follow the same trend as the calculations according to Eq.~(\ref{eq:TotTransmJoint}). The excellent agreement in Fig.~\ref{fig:FigTransmission} between the measurements and the calculations shows that any additional absorption in the laser-modified zones can be safely neglected. This is consistent with the reduced heat-affected zone produced by ultrashort pulses.\\

\begin{figure}
\centering
\includegraphics[width=\linewidth]{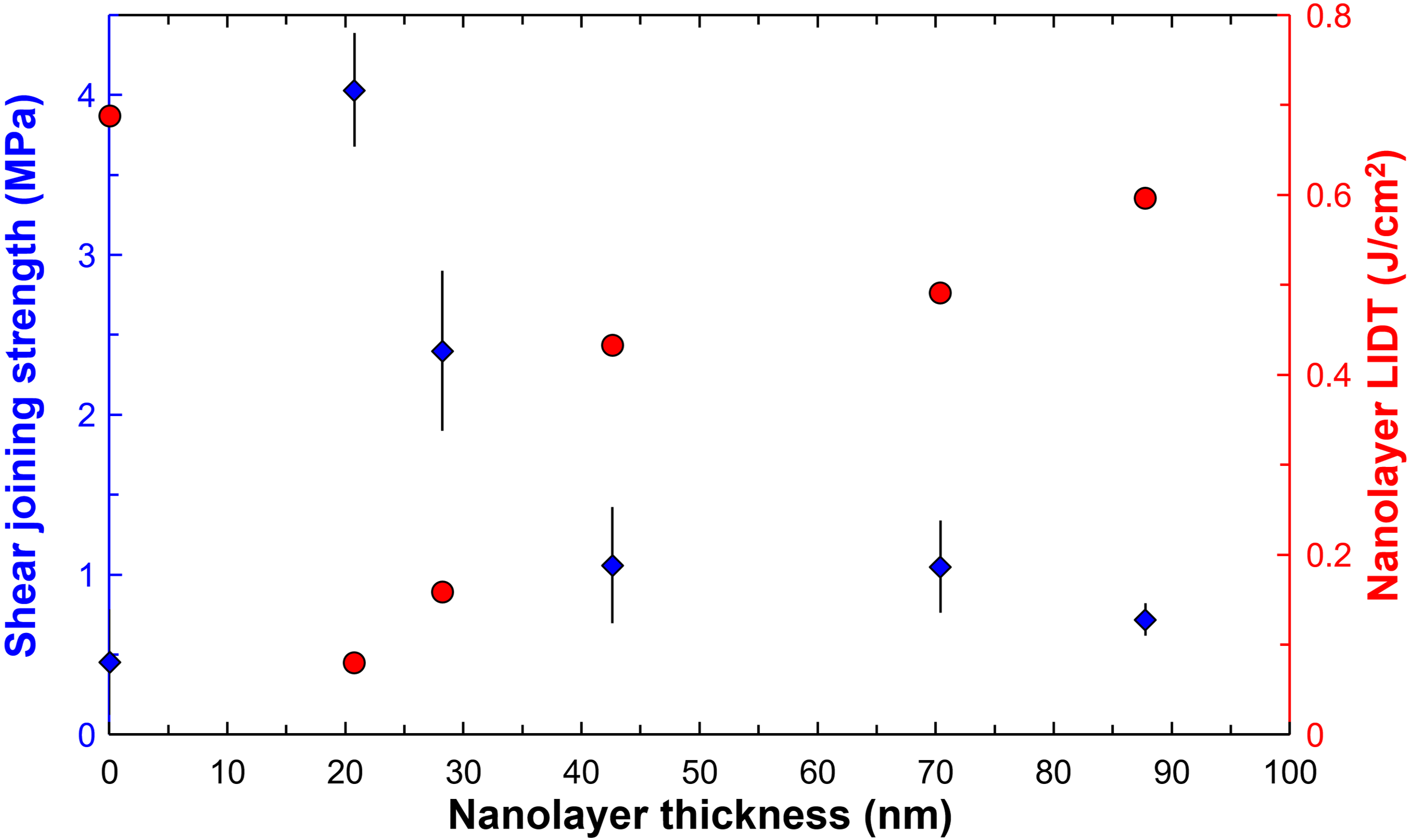}
\caption{\label{fig:FigLIDTbreaking} Dependence of the shear joining strength (blue) and the laser-induced damage threshold (LIDT) of the gold nanolayer (red) as a function of its thickness. The shear joining strength is averaged over five welds, and the error bars stand for the standard deviation.}
\end{figure}

Let us now determine the gold nanolayer thickness for which ultrashort laser welding is the most efficient. The corresponding results are displayed in Fig.~\ref{fig:FigLIDTbreaking} (blue points). Without gold nanolayer, the shear joining strength is measured to be $\approx 450$~kPa. This modest value corresponds to that obtained for samples in optical contact, without laser irradiation. One can thus conclude that the laser irradiation has no effect, in good agreement with the unmodified interface in Fig.~\ref{fig:FigIRmicroscopy}(b) compared to Figs.~\ref{fig:FigIRmicroscopy}(c)--(g). In contrast, the bonding is much more effective with a gold nanolayer. This result is independent of the thickness of the silicon sample, provided that the nonlinear focal shift is properly precompensated (see Section~S4, Supporting Information). Shear joining strengths $>4$~MPa are obtained for 21-nm nanolayers, which is comparable to the strengths obtained with adhesive bonding. The striking feature in Fig.~\ref{fig:FigLIDTbreaking} is that the thinner the nanolayer, the higher the shear joining strength. This result is not intuitive as a thicker nanolayer leads to higher absorption.\\

In order to explain this result, the laser-induced damage threshold (LIDT) has been measured for all tested nanolayers (red points in Fig.~\ref{fig:FigLIDTbreaking}). The LIDT increases with the nanolayer thickness up to a level below that of uncoated silicon, in good agreement with Refs.~\cite{Matthias1994,Stuart1996,Kruger2007}. In a very recent study, Tsibidis \textit{et al.} have shown that this behavior is attributable to the inhibition of electron diffusion in thin metallic films, causing the laser-produced hot electrons to remain confined in a restricted volume \cite{Tsibidis2022}. Therefore, upon relaxation, the thinner the nanolayer, the higher the energy transferred to the lattice by these ``trapped'' hot electrons. Consequently, damage is more likely to form for small nanolayers thicknesses, as confirmed by our experimental results in Fig.~\ref{fig:FigLIDTbreaking}. Given that the pulse energy is kept constant in all our welding experiments, the irradiation of nanolayers with lower LIDT values implies that higher energy densities are obtained. This leads to the conclusion that higher temperatures are reached for nanolayers exhibiting low LIDT, which is favorable for bonding the samples. This is confirmed in Fig.~\ref{fig:FigLIDTbreaking} by the opposite trends for the evolution of the shear joining strength and the LIDT of the nanolayer as a function of its thickness.\\

\begin{figure}
\centering
\includegraphics[width=\linewidth]{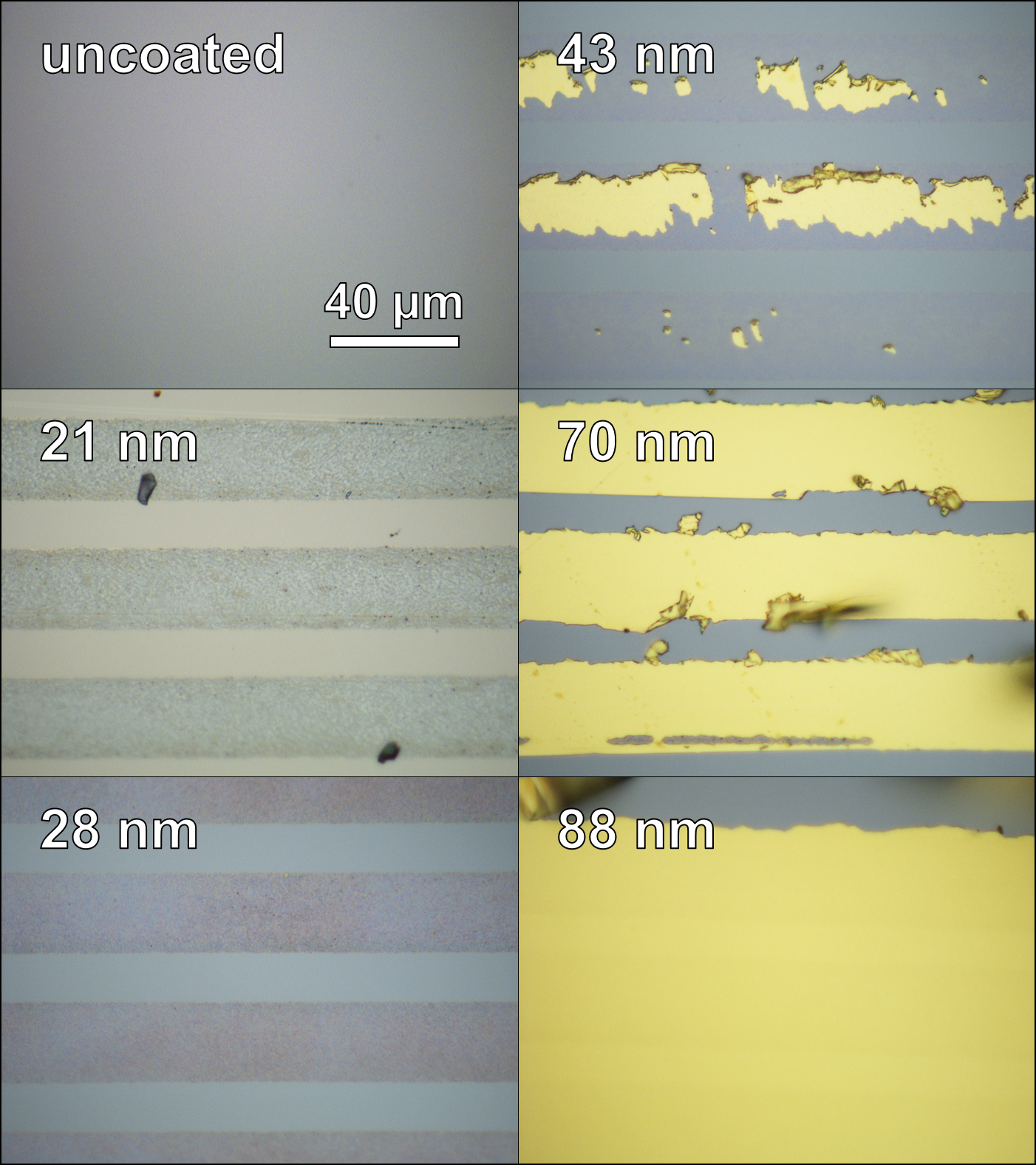}
\caption{\label{fig:FigUpperSample} Bright-field microscopy images of the exit surface of the initially uncoated top samples after sample separation for different nanolayer thicknesses deposited on the bottom samples prior to the welding. The dark, bright, and yellow features correspond to the laser-written tracks, non-irradiated areas, and gold clusters, respectively.}
\end{figure}

Ultimately, optical microscopy observations of the welds have been performed after separation. The corresponding images of the initially uncoated top samples are displayed in Fig.~\ref{fig:FigUpperSample}. Unsurprisingly, no noticeable laser-related features are observed on the exit surface of the top sample when the bottom sample is initially uncoated, which is consistent with Figs.~\ref{fig:FigIRmicroscopy}~and~\ref{fig:FigLIDTbreaking}. However, for the thinnest nanolayers ($<30$~nm), a pattern caused by the laser irradiation is unambiguously visible. This pattern is also observed for the thickest nanolayers ($>40$~nm), except that gold clusters (in yellow) are formed.\\

One may note in Fig.~\ref{fig:FigUpperSample} that the thicker the nanolayer, the larger the gold areas attached to the top sample. Two conclusions can be drawn from this result. First, for nanolayers $\ge 70$~nm, the main effect of the laser irradiation is to join the top silicon sample to gold. This is somehow analogous to our recent demonstration of semiconductor--metal ultrafast laser welding \cite{Chambonneau2021}. Second, as the bonding strength decreases with increasing nanolayer thickness (see Fig.~\ref{fig:FigLIDTbreaking}), the strength values for nanolayers $\ge 70$~nm are attributable to the adhesion of the nanolayer on the bottom sample, and not to any silicon--silicon bonds created at the interface. In contrast, for nanolayers $<30$~nm, the molten silicon in the top and bottom samples can interpenetrate, and create strong bonds after recrystallization, as supported by Raman spectroscopy in Section~S5, Supporting Information. Therefore, the strength values for the thinnest nanolayers in Fig.~\ref{fig:FigLIDTbreaking} are not related to their adhesion on the bottom sample.

\section{Conclusion}

To summarize, we have demonstrated ultrafast laser welding of silicon. This achievement relies on three aspects. First, the silicon samples have been maintained in optical contact during the welding. Second, the precompensation of the nonlinear focal shift leads to optimized energy deposition at the interface between the two silicon samples. Third, the deposition of a gold nanolayer at the interface prior to the welding leads to increased absorption. While introducing another material between two samples may resemble other methods such as adhesive bonding, employing metallic nanolayers ensures that the properties of silicon are not heavily altered. This is all the more confirmed by transmittance measurements which have shown that the laser-induced modifications do not cause additional optical losses. Counterintuitively, the thinnest nanolayers lead to the highest shear joining strengths ($>4$~MPa). This is in excellent agreement with the fact that the damage threshold of the nanolayer increases with its thickness, which implies that higher temperatures can be reached at the interface for a constant pulse energy. This result is especially important for envisioning future applications, as the deposition of a few tens of atomic layers of metal (van der Waals radius of gold: 166~pm) should not heavily alter the different properties of silicon. In other words, the technique we propose has to be considered more as local absorption enhancement rather than coating of a few hundred nanometers. To further optimize the silicon--silicon ultrafast laser welding technique, metallic nanolayer thicknesses $<20$~nm (produced with atomic layer deposition) will be employed. Also, different metallic nanolayers will be tested in the future. In particular, we anticipate that even higher bonding strengths can be reached with optimized metallic nanolayers.

\section*{Acknowledgements}

The authors thank D.~Schelle and C.~Otto (Friedrich Schiller University Jena, Institute of Applied Physics, Abbe Center of Photonics, Germany) for nanolayer thickness measurements and wafer cutting, respectively.

\section*{Funding}

This research has been supported by the Bundesministerium für Bildung und Forschung (BMBF) through the glass2met project, grant No. 13N15290.

\section*{Conflict of interest}

The authors declare no conflict of interest.

\bibliographystyle{unsrt}
\bibliography{refs}

\end{document}